\documentclass[12pt]{article}
\usepackage{graphicx}
\usepackage{amsmath}
\usepackage{longtable}

\begin{document}

\makeatletter
\renewcommand*{\@cite}[2]{{#2}}
\renewcommand*{\@biblabel}[1]{#1.\hfill}
\makeatother

\title{Extinction Law at a Distance up to 25 kpc toward the Galactic Poles}
\author{G.~A.~Gontcharov\thanks{E-mail: georgegontcharov@yahoo.com}}

\maketitle

Pulkovo Astronomical Observatory, Russian Academy of Sciences, Pul\-kov\-skoe sh. 65, St. Petersburg, 196140 Russia


Photometry from the Tycho-2, 2MASS, and WISE catalogues for clump and branch giants at
a distance up to 25 kpc toward the Galactic poles has allowed the variations of various characteristics of
the infrared interstellar extinction law with distance to be analyzed. The results obtained by the extinction
law extrapolation method are consistent for different classes of stars and different characteristics as well
as with previous studies. The conventional extinction law with a low infrared extinction is characteristic
of only a thin layer no farther than 100 pc from the Galactic plane and of two thin layers near $Z=-600$
and $+500$ pc. Far from the Galactic plane, in the Galactic halo, the infrared extinction law is different: the
extinction in the $Ks$, $W1$, $W2$, $W3$, and $W4$ bands is, respectively, $0.17$, $0.16$, $0.16$, $0.07$,
and $0.03$ of the
extinction in the $V$ band. The accuracy of these coefficients is $0.03$. If the extinction law reflects primarily
the grain size distribution, then the fraction of large dust grains far from the Galactic plane is greater than
that in the circumsolar interstellar medium.

\newpage

\section*{INTRODUCTION}

The wavelength dependence of interstellar extinction
(extinction law) had long seemed constant in
space and time and was expressed by one extinction
coefficient $R_V\equiv A_V/E_{(B-V)}\approx3.1$. Variations
of the extinction law were admitted only in some
dense clouds and star-forming regions. The data on
stars and the interstellar medium predominantly no
farther than 100 pc from the Galactic plane served
as a basis for such a view. However, the Tycho-2
(H\o g et al. 2000), 2MASS (Skrutskie et al. 2006),
WISE (Wright et al. 2010), and other catalogues with
accurate multiband photometry for faint stars over the
entire sky in various wavelength ranges cover more
distance regions.

The extinction law is apparently difficult to express
by one coefficient $R_V$. Therefore, other characteristics
of the extinction law, primarily in the infrared (IR),
are also used in present-day studies.

Using multiband IR photometry, Zasowski et al.
(2009) and Gao et al. (2009) were the first to reliably
detect large-scale (over many kpc) systematic
spatial variations of the extinction law in the diffuse
medium in the inner Galactic disk relative to the Sun.
The infrared-to-visual extinction ratio was found to
decrease with increasing distance from the Galactic
center (i.e., a larger IR extinction corresponds to a
denser medium). However, they also found a flatter
extinction law in the interarm space than that in the
arms (i.e., the IR extinction increases with decreasing
density of the medium). Berlind et al. (1997) determined
the visual and near-IR ($0.3-2$ microns) extinction
law based on photometry for the galaxy IC 2163,
which is partially occulted by the galaxy NGC 2207,
and found a flat law with $R_V=5$ in the spiral arms at
$A_V\approx1^m$ and an even flatter one in the interarm space
at $A_V\approx0.5^m$.

These contradictory results are interpreted by
Wang et al. (2013) as the existence of three types
of environment in the Galaxy: dense clouds with
relatively coarse dust and large IR extinction, a diffuse
environment of spirals with fine dust and small IR
extinction, and a translucent environment between
the arms with the same extinction law as that for
dense clouds. The similarity of the first and third
environments was emphasized by Wang et al. (2013):
the temperature is low, the gas is molecular, and the
charged particles are few. The coalescence of dust
grains apparently dominates over their fragmentation
in the clouds due to their high density, the fragmentation
dominates over the coalescence in the spiral
arms due to their high temperature, and there are
no grain-fragmenting processes in the translucent
environment itself.

Gontcharov (2013a) and Schultheis et al. (2015)
confirmed these trends and the results of Zasowski
et al. (2009) and Gao et al. (2009) using different
photometric data for the Galaxy, while Hutton
et al. (2015) found the same for the galaxy M~82
($R_V$ initially falls with increasing distance from the
Galactic center and then grows over many kpc).
Gontcharov (2013b) found the IR extinction law
outside the Galactic plane, at distances $550<|Z|<3100$ pc from it ($Z$ is the coordinate toward
the north Galactic pole), to be flatter than that near
the Galactic plane. Using Tycho-2 and 2MASS
photometry, Gontcharov (2012a) found $R_V$ to vary
from 2.2 to 4.4 within the kiloparsec nearest to the
Sun.

As a result of these studies, a dichotomy of the
Galactic dust medium is hypothesized. According
to this hypothesis, a thin layer of the medium in the
spiral arms is the only region in the Galaxy where
the interstellar extinction is essentially selective, i.e.,
the ratio of long-wavelength extinction to shortwavelength
one is minimal. This region also includes
the Galactic solar neighborhoods. By studying
mostly them, we attributed an essentially selective
extinction to the entire Galactic interstellar medium.
However, in the remaining regions of the Galaxy, i.e.,
at the center, on the disk periphery, and far from
the Galactic plane, the extinction is less selective,
i.e., the ratio of long-wavelength extinction to shortwavelength
one is, on average, larger than that in the
layer near the Galactic plane.

If the extinction law reflects primarily the grain size
distribution, then the mentioned studies and hypothesis
suggest that there is more fine dust in a thin layer
near the Galactic plane, while there is more coarse
dust at the Galactic center, on the disk periphery, and
far from the Galactic plane.

The main goal of this study is to analyze the spatial
variations of the extinction law outside the Galactic
disk at various distances from the Galactic plane up
to the maximum possible ones using different photometric
data for different samples of stars. The study
is limited only to the directions toward the Galactic
poles.

\section*{ORIGINAL DATA}

Clump and branch giants, as high-luminosity
stars represented abundantly at high Galactic latitudes
at great distances from the Sun, are used
in this study. The former are stars with nuclear
reactions of helium conversion into carbon in the
stellar core. The latter are stars with nuclear reactions
of hydrogen conversion into helium in the shell above
the helium core (irrespective of whether helium is
converted into carbon in the core, i.e., both branch
giants and asymptotic branch giants, because their
spectral energy distributions are quite similar). Since
some of the branch giants have the same absolute
magnitude and temperature as do the clump giants,
the most luminous and red part of the branch is meant
by the branch giants in this study to separate them.

In this study, just as in all the mentioned studies
of the large-scale spatial variations of the extinction
law, we apply the extinction law extrapolation
method alternatively called the color ratio method.
It was described, for example, by Straizys (1977)
and Majewski et al. (2011). As was shown by
Gontcharov (2012a), a complete or almost complete
sample of a large number of stars with a similar spectral
energy distribution distributed fairly uniformly in
space is needed for this method. The reddening and
extinction in the region of space under consideration
must be great compared to the photometric accuracy.
Clump and branch giants are suitable for this method,
but they must be considered separately, because their
spectral energy distributions are quite different.

Any samples from modern catalogues are at best
\emph{almost} complete, because they are devoid of (1) components
of unresolved binary systems and (2) variable
stars with the sought-for \emph{average} spectral energy
distribution. Such variables also account for some
fraction of red giants. In an unresolved binary system,
the total spectral energy distribution differs significantly
from the distribution for each of the components
if they have approximately equal magnitudes
but significantly different colors. B- and A-type stars
are such pairs for red giants, but both components
of the pair then have the same young age of $10^8$ yr
and a mass greater than 2 $M_{\odot}$. There are few such
stars, especially far from the Galactic plane. However,
as Gontcharov et al (2011) showed, the hot subdwarfs
located among the B stars on the Hertzsprung--Russell
diagram can have any age in the range $1-12$ Gyr and
any mass greater than 0.5 $M_{\odot}$. The number
of such stars in the region of space under consideration
can be estimated from above only roughly
by noting that there is no hot subdwarf within 5 pc of
the Sun, where all stars (56) are apparently known.
Therefore, it can be assumed that no more than 2\%
of the red giants did not fall into the samples under
consideration, hiding near hot subdwarfs.

The Tycho-2, 2MASS, and WISE (the new All-WISE version published in 2013) catalogues being
used are complete toward the Galactic poles and have
a photometric accuracy better than $0.07^m$ (the median
accuracy is about $0.02^{m}$) in the range of magnitudes
given in Table 1.

A serious problem of modern astronomy, which,
in particular, hinders the establishment of accurate
calibrations for astrophysical quantities, is that accurate
trigonometric parallaxes are known predominantly
for bright stars, for which there is no accurate
IR photometry. For example, it can be seen
from Table 1 that the WISE catalogue has accurate
photometry for nearby and, hence, bright stars only
in the $W3$ and $W4$ bands, though less accurate but
still acceptable photometry with an accuracy of $0.15^m$
in the $W1$ and $W2$ bands for many bright stars allows
acceptable results to be obtained in the range
of distances given in parentheses in Table 1. One
method of solving the problem with IR photometry
for bright stars was proposed and implemented
by Gontcharov (2011). The photometry for branch
giants in the $Ks$ band from the 2MASS catalogue
was calibrated using their photometry at 12 microns from
the IRAS catalogue (IRAS 1988). This gave the
Ks magnitudes for stars brighter than $5^m$ with an
accuracy of $0.18^m$. Although this is an order of magnitude
less accurate photometry than that for stars
with $5^m<Ks<14^m$, a large number of bright stars
when averaging the data allows a comparatively high
accuracy of their mean photometric characteristics to
be achieved in large cells of space. However, doubts
about the reality of the corresponding results obtain
by Gontcharov (2012à) remained. As we will see below,
these doubts will be dispelled in this study, where
the $W3$ and $W4$ magnitudes will be used instead of
the $Ks$ magnitude.

Clump and branch giants with an age older than
3 Gyr (Gontcharov 2008, 2011) and metallicity $\mathbf Z<0.008$
(for a discussion of the metallicity, see below)
prevail toward the Galactic poles. To avoid confusion,
the metallicity is designated everywhere below as $\mathbf Z$,
while one of the Galactic coordinates is designated
as $Z$. The mean absolute magnitudes $\overline{M}$ of clump
giants for such ages and their metallicities are given
in Table 1 in accordance with the PARSEC database
of theoretical evolutionary tracks and isochrones
of stars (it is also known as the Padova database,
http://stev.oapd.inaf.it/cmd; Bressan et al. 2012).
The clump of giants implies a small scatter of their
absolute magnitudes: only $\pm0.05^{m}$ about the PARSEC
calibration adopted in this study:
\begin{equation}
\label{rccalib}
M_{Ks}=-0.763-1.231(J-Ks)_0,
\end{equation}
where $(J-Ks)_0$ is the dereddened color. The above
calibration accuracy of $\pm0.05^m$ corresponds to a relative
accuracy of 2.5\% for the derived distances.

Similarly, the giants of the branch in its red part
have absolute magnitudes in the range from $M_{min}$ to
$M_{max}$ given in Table 1. The PARSEC calibration was
adopted for the branch giants:
\begin{equation}
\label{rgbcalib}
M_{Ks}=2.8-8.2(J-Ks)_0.
\end{equation}
The scatter of $M_{Ks}$ for individual stars relative to it is
$\pm1^m$. Accordingly, the relative accuracy of the derived
distances is 50\%, which is a factor of 20 poorer than
that for the clump giants.

Given the constraints on extinction toward the
poles, from $A_{B_T}<0.4^m$ to $A_{Ks}<0.03^m$, it follows
from these data that almost complete samples of
stars with accurate photometry can be obtained in
the ranges of distances from $|Z_{min}|$ to $|Z_{max}|$ listed
in Table 1. However, the fraction of red dwarfs in
the sample increases with magnitude. At $Ks>12^m$
the sample of giants is very difficult to clean of the
dwarfs. Therefore, in the studies performed so far,
$|Z_{max}|$ is sometimes considerably smaller than that
given in Table 1. For example, Gontcharov (2013b)
adopted the constraint $Ks<11^m$ and, consequently,
$|Z|<3150$ pc for the sample of clump giants.

\begin{table*}[!h]
\def\baselinestretch{1}\normalsize\tiny
\caption[]{Characteristics of the photometric bands and the corresponding samples of stars from the Tycho-2, 2MASS,
and WISE (new version) catalogues:
$m_{min}$ and $m_{max}$ are the magnitudes limiting the range in which the photometry is accurate (the range
for less accurate but acceptable
photometry is given in parentheses). The extinction ratio $A_{\lambda}/A_{V}$ was taken in accordance with the
model of Weingartner and Draine (2001) at $R_V=3.1$.
}
\label{bvjhkw}
\[
\begin{tabular}{l|ccccccccc}
\hline
\noalign{\smallskip}
    Characteristic  & $B_{T}$ & $V_{T}$ & $J$ & $H$ & $Ks$ & $W1$ & $W2$ & $W3$ & $W4$ \\
\hline
\noalign{\smallskip}
$\lambda_{eff}$ ($\mu m$) & 0.425 & 0.530 & 1.233 & 1.640 & 2.152 & 3.316 & 4.561 & 10.788 & 21.909 \\
$\omega_eff$ ($\mu m$) & 0.064 & 0.102 & 0.1850 & 0.2125 & 0.240 & 0.755 & 0.860 & 6.3614 & 4.2149 \\
$A_{\lambda}/A_{V}$ & 1.34 & 1.045 & 0.289 & 0.182 & 0.118 & 0.059 & 0.030 & 0.089 & 0.024 \\
$m_{min}$ & 0 & 0 & 5 & 5 & 5 (0) & 7 (5.5) & 5 (4) & $-5$ & $-5$ \\
$m_{max}$ & 11 & 10.5 & 15 & 14.5 & 14 & 16 & 15 & 10 & 6.5 \\
$\overline{M}$ of clump & 1.8 & 0.85 & $-0.9$ & $-1.35$ & $-1.5$ & $-1.5$ & $-1.5$ & $-1.5$ & $-1.5$ \\
$M_{min}$ of branch & 0 & $-2$ & $-5.5$ & $-6.3$ & $-6.5$ & $-6.5$ & $-6.5$ & $-6.5$ & $-6.5$ \\
$M_{max}$ of branch & 1.2 & $-0.5$ & $-1.7$ & $-2.5$ & $-3$ & $-3$ & $-3$ & $-3$ & $-3$ \\
$|Z_{min}|$ of clump, pc & 0 & 0 & 150 & 190 & 200 (0) & 500 (250) & 200 (125) & 0 & 0 \\
$|Z_{max}|$ of clump, pc & 600 & 740 & 14400 & 14100 & 12600 & 31600 & 20000 & 2000 & 400 \\
$|Z_{min}|$ of branch, pc & 0 & 0 & 1200 & 1800 & 2000 (0) & 5000 (2500) & 2000 (1260) & 0 & 0 \\
$|Z_{max}|$ of branch, pc & 760 & 1380 & 20900 & 25000 & 25000 & 63000 & 40000 & 4000 & 800 \\
\hline
\end{tabular}
\]
\end{table*}


Since IR photometry at least in one band is needed
for a successful application of the method, we see from
Table 1 that the method is efficient toward the poles:
\begin{itemize}
\item in the range $0<|Z|<700$ pc when using the
branch giants common to Tycho-2 ($B_T$, $V_T$)
and 2MASS ($Ks$), it was implemented by Gontcharov (2012a);
\item in the same range of distances for the branch
giants common to Tycho-2 ($B_T$, $V_T$) and WISE ($W3$, $W4$), it was implemented in this study;
\item in the range $500<|Z|<5000$ pc for the clump giants common to 2MASS ($H$, $Ks$)
and WISE ($W1$, $W2$), it was implemented by Gontcharov (2013b) for the range $550<|Z|<3100$
pc ($Ks<11^m$) using photometry from the old WISE version; it was implemented
again in this study with photometry from the new WISE version at $Ks<12^m$;
\item in the range $2500<|Z|<25000$ pc for the
branch giants common to MASS ($J$, $H$, $Ks$) and WISE ($W1$, $W2$), it was implemented in
this study.
\end{itemize}

Thus, in three parts of this study, three different
samples of stars in three ranges of distances are used
to analyze the extinction law. These results are then
compared between themselves and with the extinction
law from Weingartner and Draine (2001; below
referred to as WD2001). At $R_V=3.1$ it is presented
in the $A_{\lambda}/A_{V}$ row of Table 1 (we have in mind the
standard $V$ band with $\lambda_{eff}=553$ nm, as distinct from
the $V_T$ band with $\lambda_{eff}=530$ nm). For $0.4<\lambda<1.2$ microns
this law is consistent with the law adopted in
the PARSEC database based on Cardelli et al. (1989)
and other laws at $R_V=3.1$.

$A_{\lambda}/A_{V}$ slightly depends on the characteristics of
stars, but it may be considered constant for the entire
variety of giants with an accuracy in the infrared
sufficient for us.

\section*{BRANCH GIANTS CLOSER THAN 700 pc}

Based on photometry from the Tycho-2 and
2MASS catalogues, Gontcharov (2011) selected
30 671 KIII branch giants. The photometric distance
was calculated for each of them. Toward the Galactic
poles, this must be an almost complete sample of
stars of this class up to a distance of 700 pc from the
Sun. Based on the photometry of these stars from the
same catalogues, Gontcharov (2012a) determined
the spatial variations of the extinction coefficient
\begin{equation}
\label{vkbv}
R_V=1.12E_{(V_T-Ks)}/E_{(B_T-V_T)}.
\end{equation}
In this study, for comparison with other results,
we will present some of the results from
Gontcharov (2012a) in a different form. For the same
photometric data, we will determine the $R_V$ variations
with $Z$ coordinate for a subsample of the original
30 671 branch giants falling into the cylinder with a
radius of 150 pc around the Sun elongated along the
$Z$ axis. The radius of the cylinder is chosen in such a
way that, according to Gontcharov (2012a; see Fig. 6
there), there are no significant variations of the coefficient
$R_V$ in the $X$ and $Y$ directions inside it. Such a
cylinder turns out to contain 1355 stars. The drop in
their density in the range $|Z|<700$ pc corresponds to
the barometric law, suggesting that the sample in this
range is complete. Instead of the division into regions
of space (as was done by Gontcharov (2012a)), here
we apply a moving averaging of the coefficient $R_V$
along the $Z$ coordinates over 40 points. The result
is represented by the gray solid curve in Fig. 1 (it
coincides almost everywhere with the black solid
curve and the black dashed curve that are discussed
below). This curve represents the same results as
those in Fig. 6 from Gontcharov (2012a) and in Fig. 1
from Gontcharov (2013b).

Figure 2a compares the results for the northward
(black curve) and southward (gray curve; $|Z|$ is taken
instead of $Z$) directions. The curves are seen to
show some anticorrelation: the minima of one curve
correspond to the maxima of the other curve and vice
versa. In any case, this appears as if the dust medium
northward of the Galactic plane ``knows'' about the
state of the medium at the same distance southward
of this plane. Gontcharov (2013b) was the first to
describe this effect and assumed its periodic pattern
depending on the $Z$ coordinate. This is explainable
if the dust medium at different distances from the
Galactic plane is assumed to have formed or changed
simultaneously at the same $|Z|$ northward and southward.
This can be either a manifestation of density
waves in the vertical direction or a result of symmetric
processes during the formation of the Galaxy. In
addition, this phenomenon can be an artefact that
arose during the calculations or a systematic error of
the photometry from the 2MASS catalogue, which
was observed by two-ground based telescopes from
the Earth's northern and southern hemispheres.

To test the latter assumption, let us exclude the
ground-based results: for the same branch giants, we
will use the same Tycho-2 photometry and the $W3$
and $W4$ bands from the new WISE version. Instead
of estimating the coefficient $R_V$ from Eq. (3), we will
estimate it from the formulas
\begin{equation}
\label{vw3bv}
R_V=1.07E_{(V_T-W3)}/E_{(B_T-V_T)},
\end{equation}
\begin{equation}
\label{vw4bv}
R_V=1.02E_{(V_T-W4)}/E_{(B_T-V_T)}
\end{equation}

The proportionality factors in Eqs. (4) and (5) were
chosen in such a way that the values of $R_V$ obtained
from Eqs. (3)--(5) in this study were, on average,
equal in the range $|Z|<700$ pc under consideration.
Thus, in all three cases, we have $\overline{R_V}=3.38$ and the
corresponding factors in Eqs. (4) and (5) are $1.07$
and $1.02$. They agree with those according to the
WD2001 law: $1.09$ and $1.02$. The corresponding
nonzero extinction in the W3 and W4 bands, i.e., at
$\lambda=11$ and 22 microns, is produced by silicates (Li 2005).

The derived $Z$ dependences of the quantities (4)
and (5) are indicated in Fig. 1 by the black solid curve
and the black dashed curve, respectively (a moving
averaging over 40 points was applied). The accuracy
of all results in Fig. 1 is approximately the same and
can be estimated as $\pm0.2$ in accordance with the
results from Gontcharov (2012a) (the vertical bar in
Fig. 1).

The agreement between all curves in Fig. 1 is so
good that they largely merge together. They diverge
noticeably only for $-200<Z<-50$, $-2<Z<14$, $100<Z<125$
and $280<Z<400$ pc. The accuracy
of the $W3$ and $W4$ photometry in this range
of distances is higher than that of the $Ks$ photometry,
which, as has been noted above, was calibrated
mainly based on IRAS photometry. Therefore, the
black curves are more trustworthy, though the difference
between the curves does not indicate that some
of them are erroneous: the three coefficients under
consideration reflect the extinction law for different
wavelengths.

The steep rise of the black curves at $-2<Z<14$ pc apparently
reflects an enhanced content of large
dust grains in the immediate Galactic neighborhoods
of the Solar system predominantly northward of the
Galactic plane. This result is consistent with the result
of Kr\"uger et al. (2015), who detected an enhanced
concentration of large (more than 1 microns) grains in
the interstellar dust flow entering the Solar system
precisely from the northern Galactic hemisphere together
with the flow of neutral hydrogen and helium
based on data from the dust detector onboard the
Ulysses spacecraft. Moreover, the Galactic coordinates
of the entry point of this flow into the Solar system
(near $l=3^{\circ}$, $b=21^{\circ}$) and the exit point opposite
to it (near $l=183^{\circ}$, $b=-21^{\circ}$) roughly correspond to
the regions of maximum extinction in the Gould Belt
($l=15^{\circ}$, $b=19^{\circ}$ and $l=195^{\circ}$, $b=-19^{\circ}$) detected
by Gontcharov (2009, 2012b). Thus, the investigation
of dust grains in the Solar system and the
photometry of stars in its neighborhoods consistently
show the connection between the Solar system and
the Gould Belt as a container of dust and gas.

Apart from the peaks, all curves show a minimum
near the Galactic plane at $|Z|<100$ pc, maxima at
$100<|Z|<350$ pc, minima (even deeper than
near the Galactic plane) at $Z\approx-580$ and $Z\approx+500$ pc,
and again rise at larger $|Z|$. These $R_V$ variations are
very large, from 2.1 to 4.3 (even if the peak at $Z\approx0$ is
ignored), and were discussed in detail by Gontcharov
(2012a, 2013b).

In Figs. 2b and 2c, the results northward and
southward of the Galactic planes are compared just
as in Fig. 2a but for the quantities (4) and (5), respectively.
All of the same effects are seen. Consequently,
the match between the $R_V$ variations northward and
southward of the Galactic plane is not a systematic
error of the 2MASS photometry. However, the periodicity
of the $R_V$ variations with $Z$, especially with
a half-period of 656 pc, pointed out by Gontcharov
(2013b) is imperceptible on both graphs. This periodicity
may be a mathematical artefact.

\section*{CLUMP GIANTS CLOSER THAN 5 kpc}

Gontcharov (2013b) selected 2333 and 2483
clump giants with accurate 2MASS and WISE (the
old version, the W1 and W2 bands) photometry in
fields with a radius of $8^{\circ}$ around the north and south
Galactic poles, respectively. The photometric distances
were calculated for all stars. The derived variations
of $E_{(H-W1)}/E_{(H-Ks)}$ and $E_{(H-W2)}/E_{(H-Ks)}$
in these fields with $|Z|$ (rather than the distance, as
in Gontcharov (2013b)) are indicated in Fig. 3 by the
gray curves: on the left and the right for the north and
south poles, respectively.

A new, more accurate and extensive version of the
WISE catalogue based on a larger number of observations,
AllWISE, has been published since then.
By the same method in the same fields and in the
same $|Z|$ range, we have now selected 2415 clump
giants instead of 2333 toward the north pole and 2640
instead of 2483 toward the south one. More accurate
photometry now allows us to adopt the constraint
$Ks<12^m$ instead of $Ks<11^m$ and, accordingly,
$|Z|<5000$ pc instead of $|Z|<3150$ pc. This allowed
the number of giants under consideration to be increased
to 4661 and 5108 toward the north and south
poles, respectively. In addition, instead of the calibration
$M_{Ks}=-1.52^m$, as in Gontcharov (2013b), we
will adopt the calibration (1) that takes into account
the metallicity variations.
 
The variations of $E_{(H-W1)}/E_{(H-Ks)}$ and
$E_{(H-W2)}/E_{(H-Ks)}$ with $|Z|$ averaged over 200 points
are indicated in Fig. 3 by the black curves. The
vertical bar indicates the accuracy of the result. It
barely changes with distance, because the photometric
accuracy and the sample completeness barely
change. The differences between the black and gray
curves seen in the figure are insignificant.

\section*{BRANCH GIANTS FAR FROM THE SUN}

Branch giants are more luminous than clump giants.
This allows them to be used to analyze the
extinction law at a greater distance from the Galactic
plane. However, as has been pointed out previously,
we pay the price of a factor of 20 poorer distance accuracy
for this. In addition, the stellar density is considerably
lower far from the Galactic plane. Therefore, it
makes no sense to analyze the short-period variations
of the extinction law using branch giants: they are
smoothed out by inaccurate distances. However, the
systematic variations of this law can be analyzed on
scales of several kpc.

To get a sufficient number of stars, we had to
increase the radius of the fields around the Galactic
poles used in this section to $20^{\circ}$ instead of $8^{\circ}$ in the
previous section. They turned out to contain 356 703
and 376 984 stars with $Ks<14^m$ and accurate (better
than $0.05^m$) photometry from 2MASS and the
new WISE version in five bands: $J$, $H$, $Ks$, $W1$,
and $W2$.

From among such a large number of stars, we
selected the few branch giants based on color--color
diagrams. An important feature of the infrared color--color
diagrams for the neighborhoods of the Galactic
poles is that the stars are located on or near the
line of normal colors (zero reddening), because the
extinction and reddening are very small. The observed
distribution of all field stars on the diagram was
compared with the theoretical distribution based on
PARSEC isochrones. Five-band photometry gives
45 different color--color diagrams. The analysis was
performed for all diagrams, but some of them turned
out to be particularly informative and are considered
here.

Figure 4 shows the $(J-Ks)$ -- $(H-W2)$ diagram
for the region with a radius of 6$^{\circ}$ (instead of 20$^{\circ}$ for
the convenience of demonstration) around the north
pole. For the south pole and the regions with a radius
of 20$^{\circ}$, the diagrams look the same, and all of the
patterns found are retained. The stars are indicated
by the gray cloud of points. The isochrones (in Fig. 4a
without any correction for reddening) for the stages
from red dwarfs to the tip of the giant branch are
superimposed on it: the dashed curve is for an age
of 10 Gyr and metallicity $\mathbf Z=0.0008$ (characteristic
of the boundary of the thick disk and the halo), the
thin curve is for an age of 10 Gyr and metallicity
$\mathbf Z=0.005$, and the thin smooth curve is for an age
of 6 Gyr and metallicity $\mathbf Z=0.015$ (solar metallicity,
according to the PARSEC database). Approximate
averaged stellar characteristics are given near the
isochrones. The stellar masses in solar masses are
shown near the main sequence extending from the
top to the left: 0.1, 0.3, 0.7, and 0.9. In the lower
left corner, the isochrones turn sharply when passing
from the main sequence to the stage of giants. The
corresponding classes of giants are given near the
isochrones: GIII, KIII, and MIII. The MIII stars are
the needed branch giants.

It is immediately apparent that using this and
similar $(J-H)$ -- $(H-W2)$ diagrams, we can reject
the bulk of the main sequence and the clump giants
through the condition $(J-Ks)>0.6^m$ and the condition
$(J-H)>0.5^m$ obtained similarly.

The cloud of points in Fig. 4 is shifted by interstellar
reddening. The latter can be determined.
As can be seen from the cloud position relative to
the isochrones, if the reddening is disregarded, then
most of the stars turn out to have approximately solar
metallicity and age. However, even for red dwarfs with
$M_{Ks}\approx6^m$ having $Ks\approx14^m$ and, consequently,
$|Z|\approx400$ pc, i.e., located behind the bulk of the
absorbing layer, the reddening is significant, while
the solar metallicity is atypical. According to Peng
et al. (2012) and Schlesinger et al. (2014), who
analyzed many studies of the vertical metallicity
gradient in the Galaxy, on average, $\overline{\mathbf Z}=0.008$ at
$|Z|=400$ pc. For this condition to be met, the
isochrones must be shifted so that most of the red
dwarfs are between the isochrones $\mathbf Z=0.005$ and
$0.015$. The shift is exactly the mean reddening at $Z=400$ pc
toward the north Galactic pole: $E_{(J-Ks)NGP}=0.030^m$. $E_{(J-Ks)SGP}=0.023^m$ was obtained for the
south pole by the same method. These estimates
were used in the calibrations (1) and (2). This
independent reddening estimate near the Galactic
poles completely agrees with the estimate from
Gontcharov (2012b). The isochrones shifted by the
reddening $E_{(J-Ks)NGP}=0.030^m$ are shown against
a background of the same cloud of stars in Fig. 4b.
The thickest polygonal line here indicates the boundary
adopted in this study to separate the branch
giants (to the rightward and below the line) from the
remaining stars. This line is drawn as the bisectrix
of the angle between the actual isochrones for red
dwarfs and branch giants but with allowance made for
the photometric errors and for the fact that although
there are branch giants among the stars bluer than
some limit, they are lost among the remaining stars
(on the diagram, this limit is $(J-Ks)=0.8^m$).

In the region of red dwarfs and branch giants,
the positions of the isochrones change only slightly
with age: the isochrones of the same metallicity with
an age from 6 to 12 Gyr coincide within the accuracy
of the photometry used ($0.02^m$). However, in
contrast to the age, the metallicity of the stars under
consideration is determined very accurately from
their positions relative to the isochrones. Whereas
for most comparatively close red dwarfs $0.005<\mathbf Z<0.015$, for much more distant branch giants, as can
be seen from Fig. 4b, the isochrones $\mathbf Z=0.005$ and
$0.015$ go to the right much farther than the real stars.
The corresponding characteristic, the length of the
giant branch, is a good metallicity proxy. It follows
from this characteristic that most of the branch giants
in our sample for both poles have a metallicity
in the range $0.0008<\mathbf Z<0.002$. This estimate is
consistent with other few estimates of the vertical
metallicity gradient in the Galaxy in the range of distances
$2.5<|Z|<25$ kpc under consideration
(Kordopatis et al. 2011; Posbic et al. 2012; Carrell
et al. 2012; Peng et al. 2012; Katz et al. 2011; Bilir
et al. 2012; Gontcharov 2013b). Within the error limits,
all these estimates are between the two functions
indicated in Fig. 5 by the curves: $0.0053Z^{-0.95}<\mathbf Z<0.0067Z^{-0.54}$. In this study, the distribution
of branch giants with different $|Z|$ relative to the
isochrones also allows the vertical gradient to be estimated:
it coincides with the upper curve in the figure,
i.e., $\mathbf Z=0.0067Z^{-0.54}$.

Another example of a highly informative color--color diagram, $(J-W1)$ -- $(W1-W2)$, is shown in
Fig. 6 for the north Galactic pole without any correction
for reddening (the diagram for the south pole
looks similar). The masses of dwarfs and the classes
of giants are given near the isochrones, just as in
Fig. 4. Although theoretically $E_{(J-W1)}>E_{(J-Ks)}$,
it can be seen from the figure that most of the red
dwarfs (the top of the cloud of points) lie between
the isochrones $\mathbf Z=0.005$ and 0.015. This means
that the reddening for these stars (i.e., within 400 pc
of the Sun) is $E_{(J-W1)}\approx0$. The rest of the cloud
of points (i.e., more distant stars) shows some reddening.
Exactly the same picture is also seen for
the south pole and on the $(J-W2)$ -- $(W1-W2)$
diagram. The simplest explanations of this are as follows:
(1) the WISE photometry differs systematically
from the 2MASS photometry, or (2) the isochrones
for the $W1$ and $W2$ bands are slightly erroneous, or
(3) the extinction in the range from $J$ to $W1$ is actually
``gray'', i.e., does not depend on the wavelength,
i.e., $A_J\approx A_{W1}$. The latter explanation is confirmed
below, though it does not reject the remaining ones.

The thickest polygonal line in Fig. 6 is the line
separating the branch giants (to the right and lower)
from the remaining stars.

As a result, we selected 804 and 1068 branch
giants in the entire range of distances in the regions
with a radius of 20$^{\circ}$ around the north and south poles,
respectively.

The variations of the absolute magnitudes $M_{Ks}$
with $|Z|$ are interesting for checking our sample of
branch giants. At $|Z|<25$ kpc the stars fill the
entire range $-6.5^m<M_{Ks}<-3^m$. At $|Z|>25$ kpc
the faint stars gradually disappear from the sample.
This must distort significantly the results. Therefore,
we consider them only in the range $2.5<|Z|<25$ kpc. Here, 523 and 698 stars were selected for the
north and south poles, respectively.

The variations of $E_{(H-W1)}/E_{(H-Ks)}$ and $E_{(H-W2)}/E_{(H-Ks)}$ with $|Z|$ averaged over
100 points are shown in Fig. 7 as the gray curves
(for the north and south Galactic poles, on the left
and the right, respectively). The vertical bar indicates
the accuracy of this result in the range $4<|Z|<25$ kpc. In the range $|Z|<4$ kpc, the accuracy is
slightly higher due to the larger number of stars per
unit distance. The black curves in the same figure
were taken from the previous section of this study, i.e.,
these are the variations of $E_{(H-W1)}/E_{(H-Ks)}$ and $E_{(H-W2)}/E_{(H-Ks)}$ with $Z$ derived from the clump
giants and indicated by the black curves in Fig. 3. We
see good agreement between the corresponding black
and gray curves in the common range of distances.
On average, for the two poles $E_{(H-W1)}/E_{(H-Ks)}\approx0.8$ and $E_{(H-W2)}/E_{(H-Ks)}\approx0.85$. For comparison,
the same characteristics from Zasowski et al. (2009)
and Gontcharov (2013a) near the Galactic plane
averaged over all longitudes far from the directions
to the Galactic center and anticenter are indicated
by the circles in Fig. 7. Their scatter of $\pm0.4$ is
indicated by the vertical bars. Despite the variations,
the characteristics $E_{(H-W1)}/E_{(H-Ks)}$ and $E_{(H-W2)}/E_{(H-Ks)}$ in and far from the Galactic
plane differ by more than the variation amplitude.
$E_{(H-W1)}/E_{(H-Ks)}>1.1$ and $E_{(H-W2)}/E_{(H-Ks)}>1.1$ near the Galactic plane almost everywhere, while
$E_{(H-W1)}/E_{(H-Ks)}<1.1$ and $E_{(H-W2)}/E_{(H-Ks)}<1.1$ far from it almost everywhere. This is the main
result of our study: the IR extinction anywhere far
from the Galactic plane (in fact, in the Galactic halo)
bears no resemblance to that near the plane.

Using the color excess $E_{(H-Ks)}$ everywhere as
the denominator, we apply it as a ``standard ruler''.
However, this color excess is very small toward
the Galactic poles. To check the results, we will
take a longer ruler, the color excess $E_{(J-H)}$, and
consider $E_{(H-Ks)}/E_{(J-H)}$, $E_{(H-W1)}/E_{(J-H)}$,
$E_{(H-W2)}/E_{(J-H)}$. Their variations with $|Z|$ are
shown in Fig. 8 for the north and south Galactic
poles on the right and the left, respectively. Here,
as in Fig. 7, the gray and black curves indicate
the results for branch giants and clump giants from
the previous section, respectively. As in Fig. 7,
the vertical bar indicates the accuracy of the result
in the range $4<|Z|<25$ kpc. We see good
agreement between the results for the two classes
of objects. The differences are explained by the fact
that the spectral energy distributions for the clump
and branch giants in the $J$ band differ significantly
(approximately by $0.6^m$, in contrast to the remaining
bands under consideration, where the difference is
less than $0.15^m$). Therefore, even a small admixture of
clump giants in the sample of branch giants and vice
versa distorts the average spectral energy distribution
for the sample and, consequently, the result. On average,
$E_{(H-Ks)}/E_{(J-H)}\approx0.15$, $E_{(H-W1)}/E_{(J-H)}\approx0.25$,
$E_{(H-W2)}/E_{(J-H)}\approx0.20$.

\section*{DISCUSSION}

The ``WD2001'' column in Table 2 gives the theoretical
characteristics of the extinction law according
to the WD2001 model at $R_V=3.1$. The ``$Z=0$'' column in Table 2 gives the characteristics of
the IR extinction law from Zasowski et al. (2009)
and Gontcharov (2013a) for the Galactic disk averaged
over all longitudes far from the directions to the
Galactic center and anticenter. The characteristics
are given with their standard deviations in the entire
range of distances under consideration, which exceed
appreciably the accuracies of the characteristics
for specific distances due to their long-period spatial
variations. The observed characteristics are seen to
be consistent with WD2001 within the limits of the
standard deviation. The same table shows much
greater differences between the extinction laws in the
Galactic plane (according to WD2001 or Zasowski
et al. (2009) and Gontcharov (2013a)) and far from
it according to the results of this study (in the ``large
$|Z|$'' column, where the standard deviation of the
results in the entire $|Z|$ range under consideration
is also indicated). The main result of this study
shown in Fig. 7 is seen in the ``$E_{(H-W1)}/E_{(H-Ks)}$''
and ``$E_{(H-W2)}/E_{(H-Ks)}$'' rows of the table: values
that differ by $1.5-2$ standard deviations were obtained
from stars of the same class from the same catalogues
and by the same method near and far from the Galactic
plane. The differences are so big that they leave no
doubt about their reality.

\begin{table}[!h]
\def\baselinestretch{1}\normalsize\footnotesize
\caption[]{Comparison of the characteristics of the WD2001
extinction law at $R_V=3.1$ with the results from Zasowski
et al. (2009) and Gontcharov (2013a), on average, for the
Galactic disk far from the directions to the Galactic center
and anticenter ($Z=0$) and with the results of this study,
on average, for the two Galactic poles (large $|Z|$)
}
\label{compar}
\[
\begin{tabular}{c|ccc}
\hline
\noalign{\smallskip}
    Quantity  & WD2001 & $Z=0$ & Large $|Z|$ \\
\hline
\noalign{\smallskip}
$E_{(V_T-Ks)}/E_{(B_T-V_T)}$ & 3.14 &             & 3.41$\pm0.4$ \\
$E_{(V_T-W3)}/E_{(B_T-V_T)}$ & 3.24 &             & 3.29$\pm0.4$ \\
$E_{(V_T-W4)}/E_{(B_T-V_T)}$ & 3.46 &             & 3.45$\pm0.4$ \\
$E_{(H-W1)}/E_{(H-Ks)}$      & 1.92 & 1.7$\pm0.4$ & 0.80$\pm0.2$ \\
$E_{(H-W2)}/E_{(H-Ks)}$      & 2.37 & 2.0$\pm0.4$ & 0.85$\pm0.2$ \\
$E_{(H-Ks)}/E_{(J-H)}$       & 0.60 &             & 0.15$\pm0.15$ \\
$E_{(H-W1)}/E_{(J-H)}$       & 1.15 &             & 0.25$\pm0.15$ \\
$E_{(H-W2)}/E_{(J-H)}$       & 1.42 &             & 0.2$\pm0.18$ \\  
\hline
\end{tabular}
\]
\end{table}


The IR extinction law has been investigated by
many authors. An overview of the observations
was given by Wang et al. (2014), while an overview
of the corresponding models was given by Wang
et al. (2015). Their comparison suggests the presence
of carbon grains with a radius of more than
one micron in the medium. They produce the
emission at wavelengths up to 1 mm detected by
the COBE/DIRBE, COBE/FIRAS, and Planck
telescopes. The model of Zubko et al. (2004) with
composite grains including silicates, graphite (or
amorphous carbon), an organic and water ice mantle,
and voids shows the closest correspondence to the
observations (though the WD2001 model is mentioned
more often and, therefore, is adopted below
as a standard for comparison).

However, almost all observations of the IR extinction
law refer to dense clouds near the Galactic
plane or toward the Galactic center. Here, coincident
results, within the error limits, were obtained
by Lutz (1999), Indebetouw et al. (2005), Jiang
et al. (2006), Flaherty et al. (2007), Nishiyama
et al. (2009), Fritz et al. (2011), Chen et al. (2013),
and Gao et al. (2009). The extinction law averaged
over them as a function of $1/\lambda$ for the $W4$, $W3$,
Spitzer/IRAC 8 microns, Spitzer/IRAC 5.8 microns, $W2$
(or Spitzer/IRAC 4.5 microns), $W1$ (or Spitzer/IRAC
3.6 microns), $Ks$, and $H$ bands is indicated by the circles
in Fig. 9. For comparison, the WD2001 laws at
$R_V=3.1$ and 5.5 are indicated by the lower and upper
dashed curves, respectively. The observations are
seen to fall nicely on the WD2001 law at $R_V=5.5$.
The same IR extinction law, but with a peak at
4.5 microns, was obtained by Wang et al. (2013) in the
Coalsack nebula and by Gao et al. (2013) in the
dense medium of the Large Magellanic Cloud. These
extinction peaks at 4.5 microns are indicated in Fig. 9 by
the gray and black triangles, respectively.

Not all studies at high latitudes actually refer to
the medium far from the Galactic plane. As an example
of confusion, it is worth noting the paper by
Larson and Whittet (2005), where the extinction law
is investigated at high Galactic latitudes but within
100 pc of the Sun. Obviously, their results refer to
the well-studied equatorial layer of dust with mean
$R_V=3.1$. To avoid confusion, ``far from the Galactic
plane'', or ``definitely outside the Galactic equatorial
layer of dust'', or ``in the Galactic halo'' should be
written instead of the term ``high latitudes''.

At present, the results of Davenport et al. (2014)
are apparently the only observations of the IR extinction
law that refer to the diffuse medium both near
and far from the Galactic plane (at $Z\sim1$ kpc): the
results for $b>50^{\circ}$ and $0^{\circ}<b<25^{\circ}$, i.e., near the Galactic
plane, are indicated in Fig. 9 by the open diamonds
and open squares, respectively. The closeness of the
circles and open diamonds in the figure confirms the
previously suggested hypothesis about the similarity
of the extinction laws far from the Galactic plane,
in dense clouds near this plane, and at the Galactic
center.

Note that Davenport et al. (2014) found an especially
dramatic increase in extinction (relative to $A_V$)
in the $H$ and $Ks$ bands at latitudes $25^{\circ}<b<50^{\circ}$. At
latitudes $50^{\circ}<b<90^{\circ}$, the relative extinction in $H$
and $Ks$ decreases, but, at the same time, the extinction
in $W1$ and $W2$ increases dramatically. This study
shows the same trend. This appears as a gradual
increase of the mean grain size with distance from the
Galactic plane. Davenport et al. (2014) also found
the longitude dependence of the extinction law: it is
flatter at longitudes near $90^{\circ}$, where the Local Spiral
Arm passes (but there are no studies at a longitude of
$270^{\circ}$). A detailed analysis of the variations in extinction
law with longitude and distance $|Z|$ is needed.

All of the extinction laws mentioned here are relative,
with the zero point, i.e., the extinction in some
band taken as the true one, being different in each
study. To compare the results, given that most of
them in the IR are consistent with the WD2001 extinction
law at $R_V=5.5$, we will take this value as the
zero point. Then, $A_J=0.289A_V$, $A_H=0.182A_V$, $A_{Ks}=0.118A_V$.

Three versions of the extinction law found in this
study are possible, depending on precisely which of
these estimates are used as the zero point. They
are indicated in Fig. 9 by the filled diamonds and
solid polygonal lines. In this case, the data from the
right column of Table 2 are used.
From
$E_{(V_T-W3)}/E_{(B_T-V_T)}=3.29$ and $E_{(V_T-W4)}/E_{(B_T-V_T)}=3.45$
we estimated $A_{W3}=0.074A_V$ and $A_{W4}=0.027A_V$
(the two leftmost diamonds in Fig. 9) relative to the fixed $A_{B_T}=1.34A_V$ è $A_{V_T}=1.045A_V$;
from
$E_{(H-Ks)}/E_{(J-H)}=0.15$, $E_{(H-W1)}/E_{(J-H)}=0.25$ and $E_{(H-W2)}/E_{(J-H)}=0.2$ we estimated
$A_{Ks}=0.166A_V$, $A_{W1}=0.155A_V$ è $A_{W2}=0.161A_V$
(the middle of the three curves) relative to the fixed
$A_J=0.289A_V$ and $A_H=0.182A_V$;
from
$E_{(H-W1)}/E_{(H-Ks)}=0.8$ and $E_{(H-W2)}/E_{(H-Ks)}=0.85$ we
estimated $A_{W1}=0.169A_V$ and $A_{W2}=0.168A_V$
(the upper of the three curves) relative to the fixed
$A_H=0.182A_V$ and the estimated $A_{Ks}=0.166A_V$;
from
$E_{(H-W1)}/E_{(H-Ks)}=0.8$ and $E_{(H-W2)}/E_{(H-Ks)}=0.85$ we estimated $A_{W1}=0.131A_V$ and $A_{W2}=0.128A_V$
(the lower of the three curves) relative to the fixed $A_H=0.182A_V$ and $A_{Ks}=0.118A_V$.
The accuracy of these estimates is $0.03A_V$.

Good agreement between the upper and lower
curves suggests that all of the extinctions in the $J$,
$H$, $Ks$, $W1$, and $W2$ bands correspond to one another
within one extinction law. The only difference
between the lower curve and the two remaining ones
consist in its vertical shift due to the different extinction
$A_{Ks}$ taken as the zero point. Clearly, an
uncertainty in the zero point of the extinction law
is the main problem in all such studies. It can be
solved only by using homogeneous and highly accurate
spectrophotometry in a very wide wavelength
range, from the ultraviolet to the far infrared.

An independent determination of $A_{Ks}$ is of particular
interest. Therefore, although all three versions of
the derived extinction law are equivalent, as a unified
result of the study we will take the mean between
the upper and middle curves, i.e., $A_{Ks}=0.166A_V$, $A_{W1}=0.162A_V$, $A_{W2}=0.164A_V$,
$A_{W3}=0.074A_V$, $A_{W4}=0.027A_V$. Thus, given $A_H=0.182A_V$,
we have a perfectly flat extinction law in the bands
from $H$ to $W2$, i.e., in the wavelength range from 1.4
to 5.4 microns, but the estimated extinction in the $W1$
and $W2$ bands exceeds the results by other authors by more
than the declared errors. However, the results of this
study were obtained at a much greater distance from
the Galactic plane.

\section*{CONCLUSIONS}

In this study, we receded from the Galactic plane
to $|Z|=25$ kpc toward the north and south Galactic
poles. In doing so, we detected a conventional
extinction law for the solar neighborhood with a
low IR extinction only near the Galactic plane at
$|Z|<100$ pc and in two thin layers at $Z\approx-600$ pc and $Z\approx500$ pc.
In the rest of the vertical column
with a length of 50 kpc through the entire Galaxy
and the circumgalactic space, the extinction is less
selective, i.e., the ratio of long-wavelength extinction
to short-wavelength one is greater than that in
the solar neighborhood. In the wavelength range
from 1.4 to 5.4 microns (the $Ks$, $W1$, and $W2$ bands),
the extinction law is flat, the extinction is ``gray'',
i.e., does not depend on the wavelength. We estimated
$A_{Ks}=0.166A_V$, $A_{W1}=0.162A_V$, $A_{W2}=0.164A_V$,
$A_{W3}=0.074A_V$, $A_{W4}=0.027A_V$
with an accuracy of $0.03A_V$.

If the extinction law depends primarily on the grain
size, then this study points to an increase in the
fraction of coarse dust as one recedes from theGalactic
plane and to the preservation of this state of the
medium in the entire Galactic halo at least up to
$|Z|<25$ kpc.

We also detected an increase in the fraction of
coarse dust in the immediate neighborhoods of the
Solar system at $-2<Z<14$ pc consistent with
the characteristics of the flow of coarse dust entering
the Solar system apparently from the most dusty
regions of the Gould Belt.

The variety of the Galactic dust medium could not
be detected without highly accurate (at a $0.01^m$ level)
mid-IR photometry, i.e., before the appearance of the
WISE catalogue.

\section*{ACKNOWLEDGMENTS}

In this study, we used results from the Hipparcos–
Tycho, Two Micron All Sky Survey (2MASS), and
Wide-field Infrared Survey Explorer (WISE) projects
as well as resources from the Strasbourg Astronomical
Data Center (Centre de Donnees astronomiques
de Strasbourg). The study was financially supported
by the ``Transient and Explosive Processes in Astrophysics''
Program P-7 of the Presidium of the Russian
Academy of Sciences.

\newpage

\begin{figure}
\includegraphics{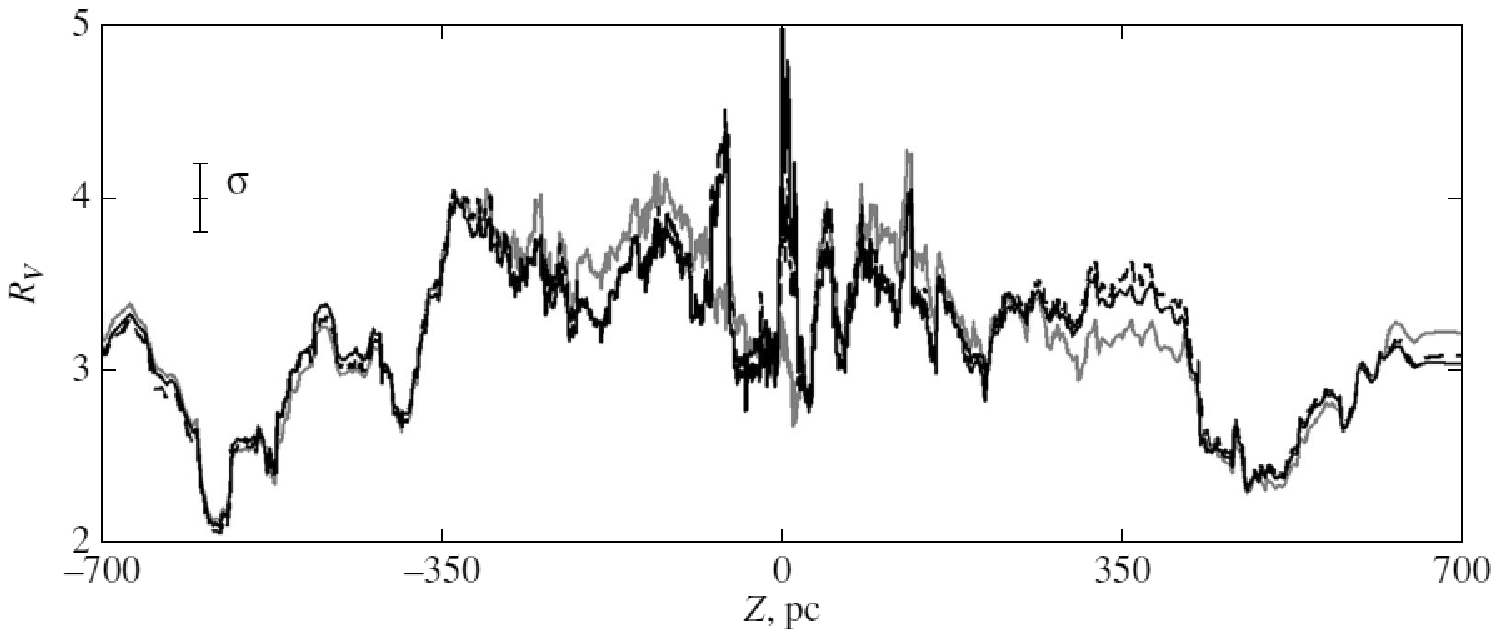}
\caption{Variations of the quantities (3) (gray curve), (4) (black dashed curve), and (5) (black solid curve)
from the photometry
of 1355 branch giants in the spatial cylinder along the $Z$ axis with a radius of 150 pc around the Sun.
}
\label{fignearrv}
\end{figure}

\begin{figure}
\includegraphics{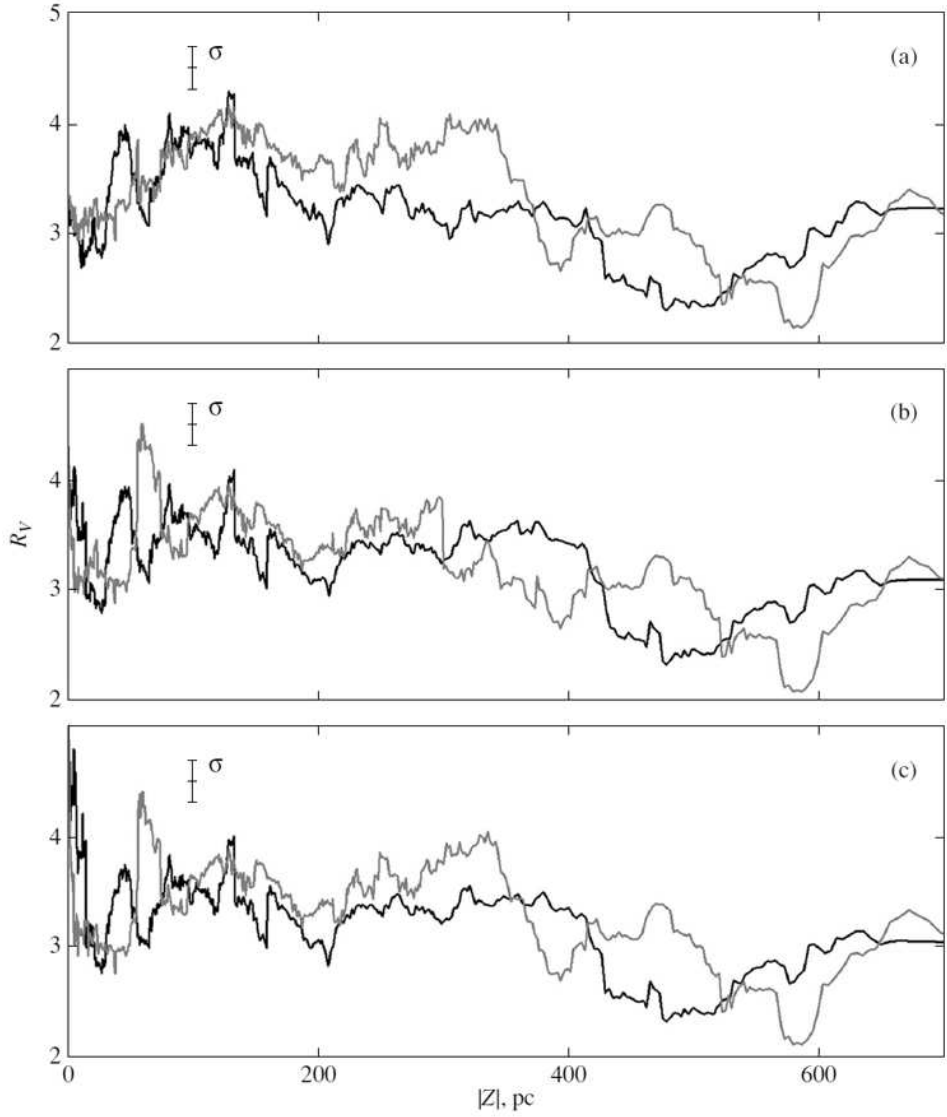}
\caption{Variations of the quantities (a) (3), (b) (4), and (c) (5) northward (black curves) and southward (gray curves)
along $|Z|$ from the photometry of 1355 branch giants in the spatial cylinder along the $Z$ axis with a radius of
150 pc around the Sun.
}
\label{fignearns}
\end{figure}

\begin{figure}
\includegraphics{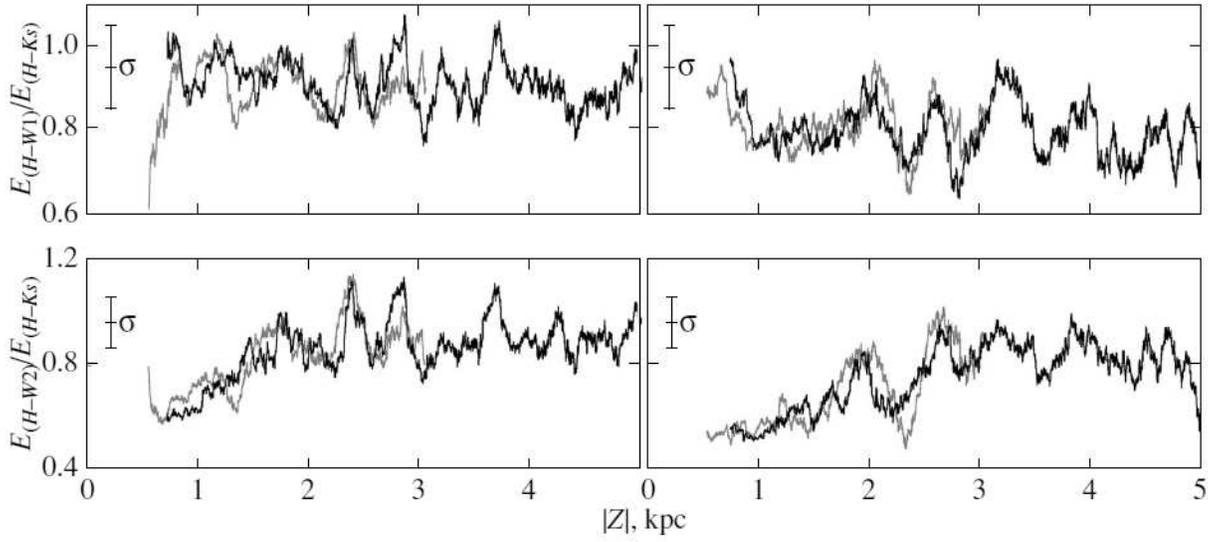}
\caption{Variations of $E_{(H-W1)}/E_{(H-Ks)}$ and $E_{(H-W2)}/E_{(H-Ks)}$ toward the north (left) and south (right)
Galactic poles from
the data for clump giants from 2MASS and the old (gray curves) or new (black curves) WISE version.
}
\label{figpaper12}
\end{figure}

\begin{figure}
\includegraphics{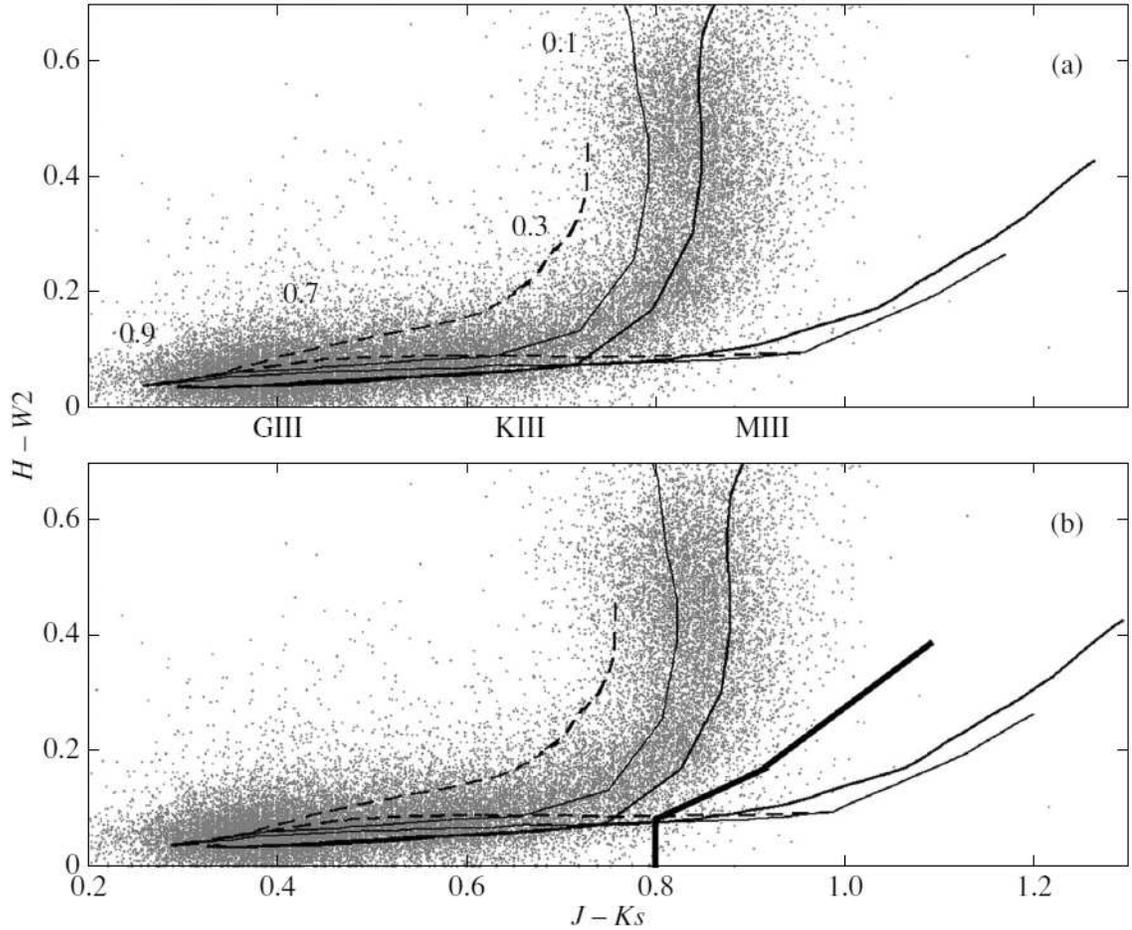}
\caption{$(J-Ks)$ -- $(H-W2)$ diagram: common 2MASS and WISE stars with accurate photometry within $6^{\circ}$ of the
north Galactic pole (the gray cloud of points); isochrones (a) uncorrected for reddening and (b) corrected for reddening
$E_{(J-Ks)}=0.03^m$ for an age of $10^{10}$ yr and metallicity $\mathbf Z=0.0008$ (dashed line), $10^{10}$
yr and $\mathbf Z=0.005$ (thin line), and
$6\times10^{9}$ yr and $\mathbf Z=0.015$ (thick smooth line). Approximate masses of dwarfs and classes of giants are shown along the
isochrones. The thickest polygonal line is the star selection line.
}
\label{jkhw2}
\end{figure}

\begin{figure}
\includegraphics{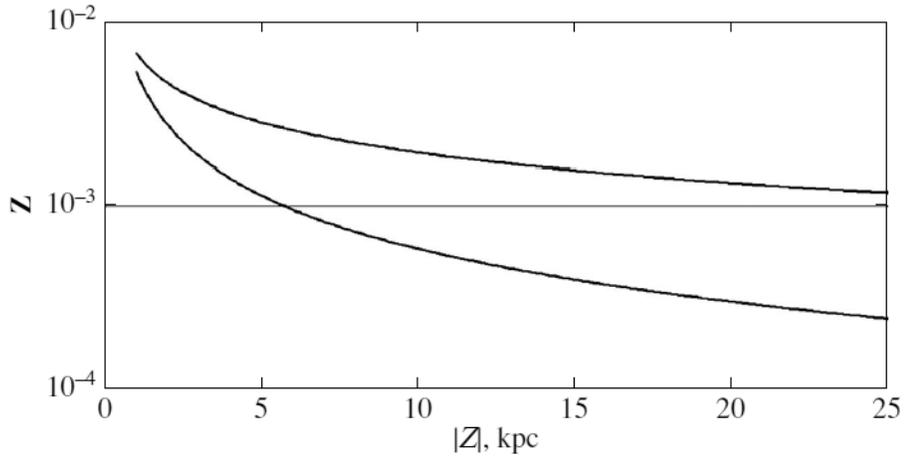}
\caption{Mean stellar metallicity $\mathbf Z$ versus $|Z|$ coordinate: the curves for the minimum and maximum estimates.
}
\label{zz}
\end{figure}

\begin{figure}
\includegraphics{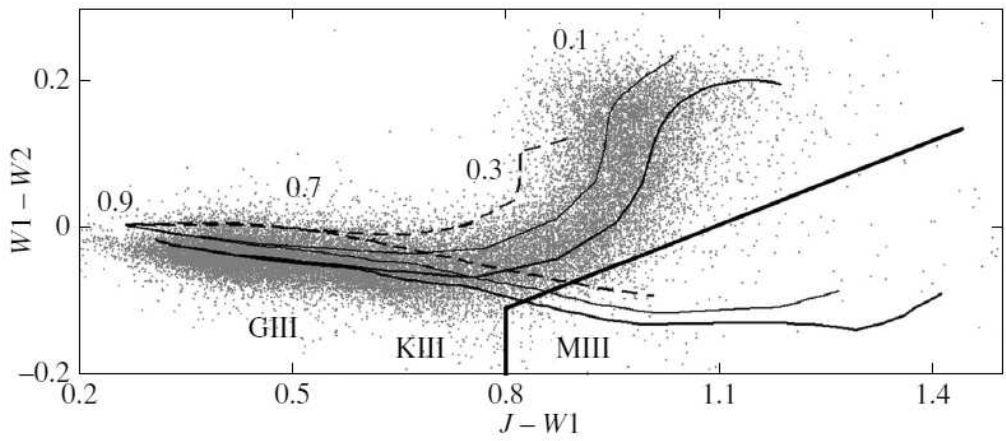}
\caption{Same as Fig. 4 for the $(J-W1)$ -- $(W1-W2)$ diagram.
}
\label{jw1w1w2}
\end{figure}

\begin{figure}
\includegraphics{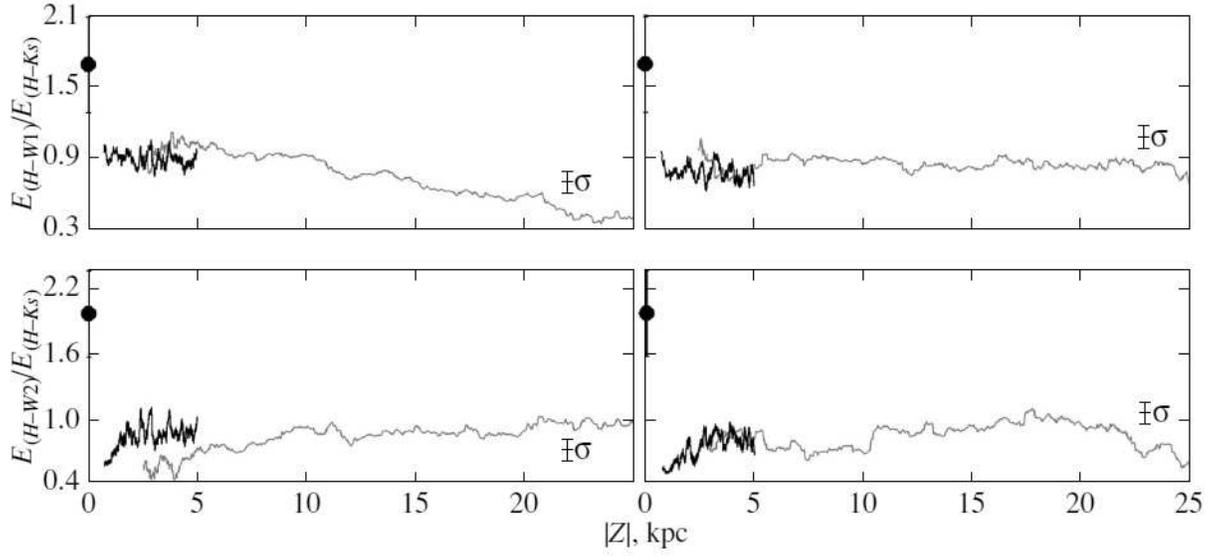}
\caption{Variations of $E_{(H-W1)}/E_{(H-Ks)}$ and $E_{(H-W2)}/E_{(H-Ks)}$ toward the north (left) and south (right)
Galactic poles
from the data for branch giants from 2MASS and the new WISE version (gray curves) in comparison with the analogous
results on clump giants from 2MASS and the new WISE version (the black curves taken from Fig. 3). The circles indicate the
characteristics from Zasowski et al. (2009) and Gontcharov (2013a) near the Galactic plane averaged over all longitudes far
from the directions to the Galactic center and anticenter. Their scatter of $\pm0.4$ is indicated by the vertical bars.
}
\label{figfar}
\end{figure}

\begin{figure}
\includegraphics{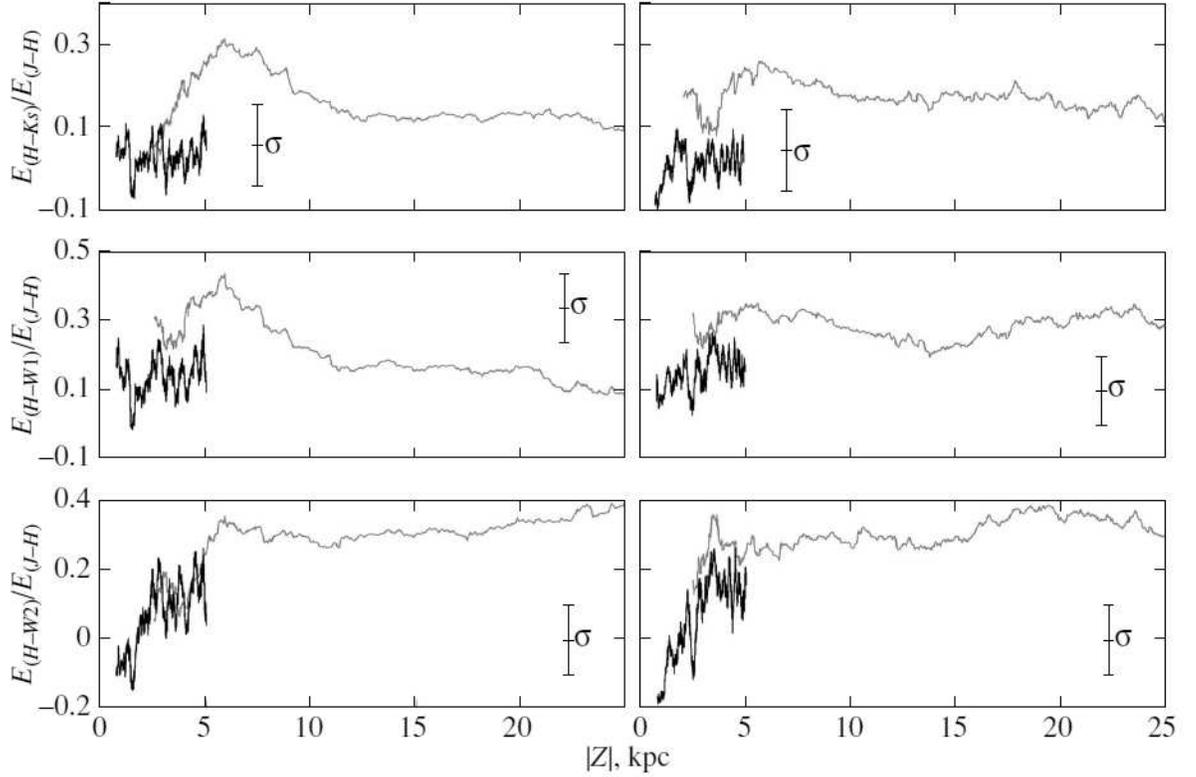}
\caption{Variations of $E_{(H-Ks)}/E_{(J-H)}$, $E_{(H-W1)}/E_{(J-H)}$ and $E_{(H-W2)}/E_{(J-H)}$ toward the north
(left) and south (right)
Galactic poles from the data for branch giants from 2MASS and the new WISE version (gray curves) in comparison with the
analogous results on clump giants from 2MASS and the new WISE version (gray curves).
}
\label{figjh}
\end{figure}

\begin{figure}
\includegraphics{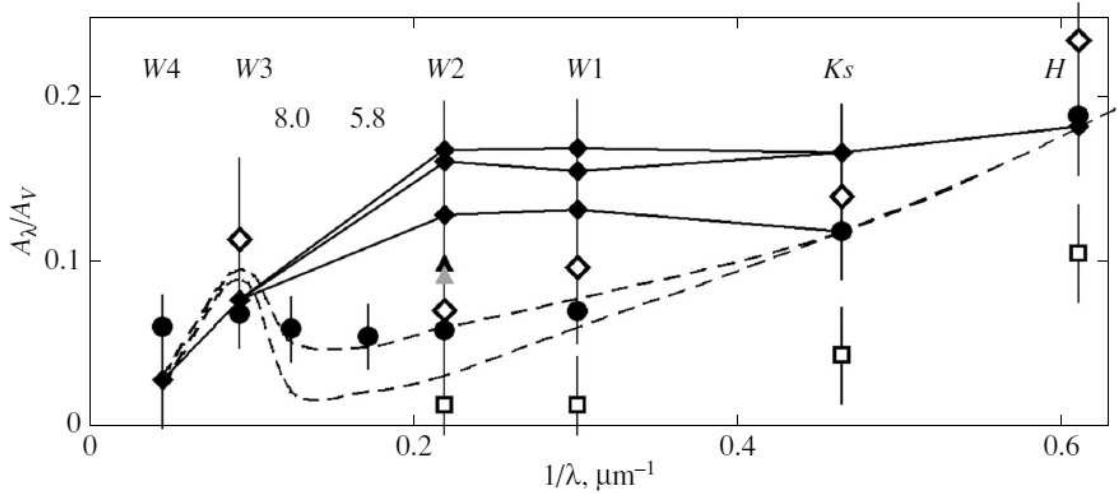}
\caption{$A_\lambda/A_V$ versus $1/\lambda$. The WD2001 extinction law
at $R_V=3.1$ and 5.5 is indicated by the gray
dashed curves, upper
and lower, respectively. The averaged results of the studies
listed in the text are indicated by the circles.
The gray and black
triangles mark the extinction peaks at 4.5 microns found by
Wang et al. (2013) and Gao et al. 2013).
The results from Davenport
et al. (2014) for $b>50^{\circ}$ and $0^{\circ}<b<25^{\circ}$ are indicated by the open diamonds and
open squares, respectively. Three versions of
the extinction law obtained in this study are indicated by the diamonds and black curves:
the upper one relative to the fixed
$A_H=0.182^m$ and estimated $A_{Ks}=0.166^m$, the middle one relative
to the fixed $A_J=0.289^m$ and $A_H=0.182^m$, and the
lower one relative to the fixed $A_H=0.182^m$ and $A_{Ks}=0.118^m$. The spectral bands are specified.
}
\label{law}
\end{figure}

\end{document}